\documentclass{iac}
\usepackage{soul}
\usepackage{fontspec}  
\usepackage{xurl}
\usepackage{tabularx}
\usepackage{booktabs}

\begin{document}

\IACconference{76}
\IAClocation{Sydney, Australia}
\IACdates{29 Sep-3 Oct}
\IACyear{2025}
\IACpapernumber{A4,1,6,x95796}
\IACcopyright{2025}{the International Astronautical Federation (IAF)}

\title{The VLA and High-Frequency SETI: Expanding the Search for
Life}

\IACauthor{Talon Myburgh*\textsuperscript{e,}}
{Centre for Radio Astronomy Techniques and Technologies (RATT), Department of Physics and Electronics, Rhodes University, Makhanda, 6140, South Africa}
{g14m8400@campus.ru.ac.za}
\IACauthor{Noah Stiegler}{SETI Institute, 339 Bernardo Ave, Suite 200,
Mountain View, CA 94043, USA}{noah.stiegler@berkeley.edu}
\IACauthor{Chenoa D. Tremblay}{SETI Institute, 339 Bernardo Ave, Suite 200,
Mountain View, CA 94043, USA}{ctremblay@seti.org}
\IACauthor{Joe S. Bright}{Breakthrough Listen, University of Oxford, Department of Astrophysics, Denys Wilkinson Building, Keble Road, OX1 3RH, UK}{joe.bright@physics.ox.ac.uk}
\IACauthor{Ross A. Donnachie}{Mydon Solutions (Pty) Ltd., 102 Silver Oaks, 23 Silverlea Road, Wynberg, Cape Town, South Africa, 7800}{company@mydon.co.za}

\abstract{The Commensal Open-Source Multimode Interferometer Cluster (COSMIC) runs software that searches for technologies elsewhere in the Universe (``technosignatures") using the Karl G. Jansky Very Large Array (VLA). Specifically, it searches for narrowband signals that drift in frequency over time as a result of Doppler motions. Although this is the first study of high-frequency technosignatures that has been published from the COSMIC system on the VLA, it follows closely on previous work completed by an undergraduate research intern. Within the field of view of the VLA, the software on COSMIC creates coherent beams directed toward stars that may contain exoplanets from the $Gaia$ catalogue. The recorded results follow a real-time software pipeline and are examined for technosignatures. Using a Taylor-Tree De-Dispersion algorithm to find narrow-band drifting signals, each recorded beam (coherent and incoherent) is searched for signals with a drift-rate with magnitudes up to $\pm$50Hz/s. All detections are stored as ``hits" with the relevant snippet of data stored for posterity. The purpose of this work is to extend the high-frequency search by reviewing data from February 2024 to the present. Our study examines the impact of previously implemented and novel filters to find a selection of candidate signals. At the final stage of the pipeline, our objective is to study these resultant candidate signals spatially through imaging. The observations therefore probe regions of frequency and signal parameter space that have received comparatively limited coverage in previous SETI surveys.}

\maketitle

\section*{Acronyms/Abbreviations}

\noindent Taylor tree de-dispersion (TTDD)

\noindent NSF Karl G. Jansky Very Large Array (VLA)

\noindent Search for Extraterrestrial Intelligence (SETI)

\noindent Field Programmable Gate Array (FPGA)

\noindent Radio Frequency Interference (RFI)

\noindent Signal to Noise Ratio (SNR)

\noindent National Radio Astronomy Observatory (NRAO)

\noindent Equivalent Isotropic Radiative Power (EIRP)

\section{Introduction}


SETI investigates electromagnetic emissions whose spectro-temporal characteristics differ from those expected of natural backgrounds or known astrophysical phenomena, treating such signals as potential indicators of advanced extraterrestrial intelligence. Although no evidence for the existence of advanced life beyond Earth has yet been found (e.g., \cite{smith2021radio}, \cite{ma2023deep}), it is increasingly important to exploit modern technological capabilities to conduct broader and more sensitive searches, and thereby assess the uniqueness of our own civilisation in the universe.

Due to modern astrophysical observing techniques and increased computational power, astronomers are performing all-sky surveys and making detailed maps of the Universe with increased efficiency and sensitivity. 
In the last decade, we have also come to understand that planets are common around stars. We also observe that the distribution of stars is relatively isotropic across a large scale. However, there is a question that remains about the distribution of life. Is it isotropic and prevalent, or is it a rare instance of a certain set of conditions? 

To help answer this question, we developed the Commensal Open-Source Multimode Interferometer Cluster (COSMIC), one of the first commensal Ethernet-based digital back-ends on the VLA Telescope, which started scientific operation in April 2023 \cite{hickish2019commensal}, \cite{cosmic2023}. At an incoming data rate of $1.7\,\rm{Tb}\,\rm{s}^{-1}$, COSMIC autonomously processes and searches these data in real time in one of the largest SETI experiments ever carried out. As a commensal system, COSMIC is designed to adapt to the primary observer’s frequency and setup, while simultaneously searching in real time for narrowband radio signals that drift in time and frequency in ways consistent with our understanding of radio-emitting technology (technosignatures).

By using an array of dishes (an interferometer), like the VLA, we have a choice between searching for technosignatures in beamformed data products or images. In beamforming, the phase information is used to simultaneously form multiple coherent beams, each the size of the point-spread function, toward key sources, and the signals from each antenna are summed together. For imaging, data from each pair of antennas are correlated, producing visibilities, and the Fourier transform of the visibilities is used to image the sky with the resolution of the point-spread function. 

The pipelines running on COSMIC uses beamforming to do the initial search. The search is completed in real-time on both the coherent beamformed data products and the incoherent sum (a two-dimensional stack of the signals detected with each online antenna). The targets for the coherent beamformed data are nearby stars (less than 3000\,pc) from the Gaia Space Telescope and with well-known proper motions \cite{czech2021breakthrough}. However, when no targets are found within the field of view, we form a coherent beam at the centre of the field, and then later determine if there are targets of interest through a coordinate search. Although the real-time pipeline does not search the entire field in imaging mode, we can image the data in post-processing to assess the characteristics of any signals identified in the beamformed data. 

To determine if a signal is radio frequency interference (Earth-based, satellite, or probes within our solar system) or a real technosignature with an astronomical origin, we need to develop a series of logical steps. In this paper, we describe a method for filtering signals detected by the real-time pipeline (``hits") to find potential frequency drifting technosignatures (``candidates"). In addition, the filtered candidates are phased to the centre of the coherent beam in which the hit was detected and imaged. Imaging the candidate extends the observable parameter space to include polarisation, spatial context, and increased sensitivity.   

The main receivers at the VLA operate between 1 and 50 GHz. This work focuses on upper-frequency receivers, covering 25--50\,GHz. Most technosignature research has been conducted at frequencies below 10\,GHz, mainly because fewer radio telescopes are equipped with receivers at higher frequencies. Nonetheless, a few smaller studies have been carried out above this range (e.g., \cite{Sardinia,mason2025conducting}).

With support from the world-leading Breakthrough Listen (BL) programme, our search for technosignatures adopts a relatively agnostic approach, reducing assumptions about the specific frequencies at which candidate signals might appear. The 25--50\,GHz band is particularly attractive because it is largely free of terrestrial radio interference, improving sensitivity to weak signals, and because it has only been explored a little in previous technosignature studies. Here we present the first large-scale programme to survey this range, covering 25\,GHz of bandwidth at 2\,Hz frequency resolution\footnote{We note that this bandwidth is not simultaneously processed. See Section \ref{sec:observation} for further details.}. This work builds on an undergraduate research project, which will be published in full at a later date.

\section{Observation}\label{sec:observation}
Starting in October 2023, COSMIC recorded data and searched for narrowband signals in 56-second time segments with simultaneous observations with the VLA at frequencies greater than 2\,GHz. However, for this study we focus on a frequency range of 25 to 50\,GHz (K-band [18--26.5\,GHz], Ka-band [26.5–-40\,GHz] and Q-band [40–-50\,GHz]).

Each time segment is considered by the COSMIC pipelines as an independent observation, and the system has no knowledge of one segment from another. The data are split into $\approx32\mathrm{MHz}$ sub-bands, and distributed across the compute cluster of 21 compute nodes with 2 discrete processing pipelines each. Per pipeline, each sub-band is up-channelised to $\approx2\mathrm{Hz}$ frequency resolution. At full capacity, a total bandwidth of up to $\approx1.344\mathrm{GHz}$ was processed simultaneously. The exact frequencies recorded change with each observation, as the setup is controlled by the primary observer. Every 5 to 10 minutes, the system is calibrated to compute the antennae's complex gains and correct the phase information. The total amount of time on each patch of sky is variable and is dictated by the primary observer. The phases were then calibrated using standard radio astronomy techniques. The complete hardware setup, process of calibration, and processing is explained in Tremblay et al. \cite{cosmic2023}.

\subsection{Calibration Verification}
The pipeline running on COSMIC follows standard radio astronomy gain solution calibration techniques and each calibration scan is graded based on the phase stability. However, this is challenging with high-frequency radio data because the troposphere makes the sky less transparent to incoming radiation. This presents itself as phase instabilities, making the final phases after gain calibration have a higher degree of variability. Calibration solutions are assigned a phase-stability grade between 0 and 1, where values above 0.6 correspond to less than a 40\% reduction in coherent sensitivity and typically indicate that only one or two antennas exhibit poor phase behaviour \cite{vlass2025cosmic}. Observations with grades below this threshold undergo additional diagnostic inspection of the calibration plots and beamformed data products, to determine whether the resulting science data remain suitable for technosignature searches. 

This study observes $\approx2$ million hits, representing potential narrow-band SETI signals that are detected by the real-time pipeline using \textsc{seticore}\footnote{\url{https://github.com/lacker/seticore}}. 
Each hit originates from a VLA observation with an intent stated as ``target", and is ascribed a grade for the calibration observation immediately proceeding it. Following the calibration pipeline, each hit can have one of three states: 1) has a grade $<0.6$ and may require inspection before trusting its result, 2) has a grade $\geq0.6$ and is considered stable, and 3) has no immediate proceeding calibration observation and therefore cannot be allocated a grade. 

\begin{figure*}
    \centering
    \includegraphics[width=\linewidth]{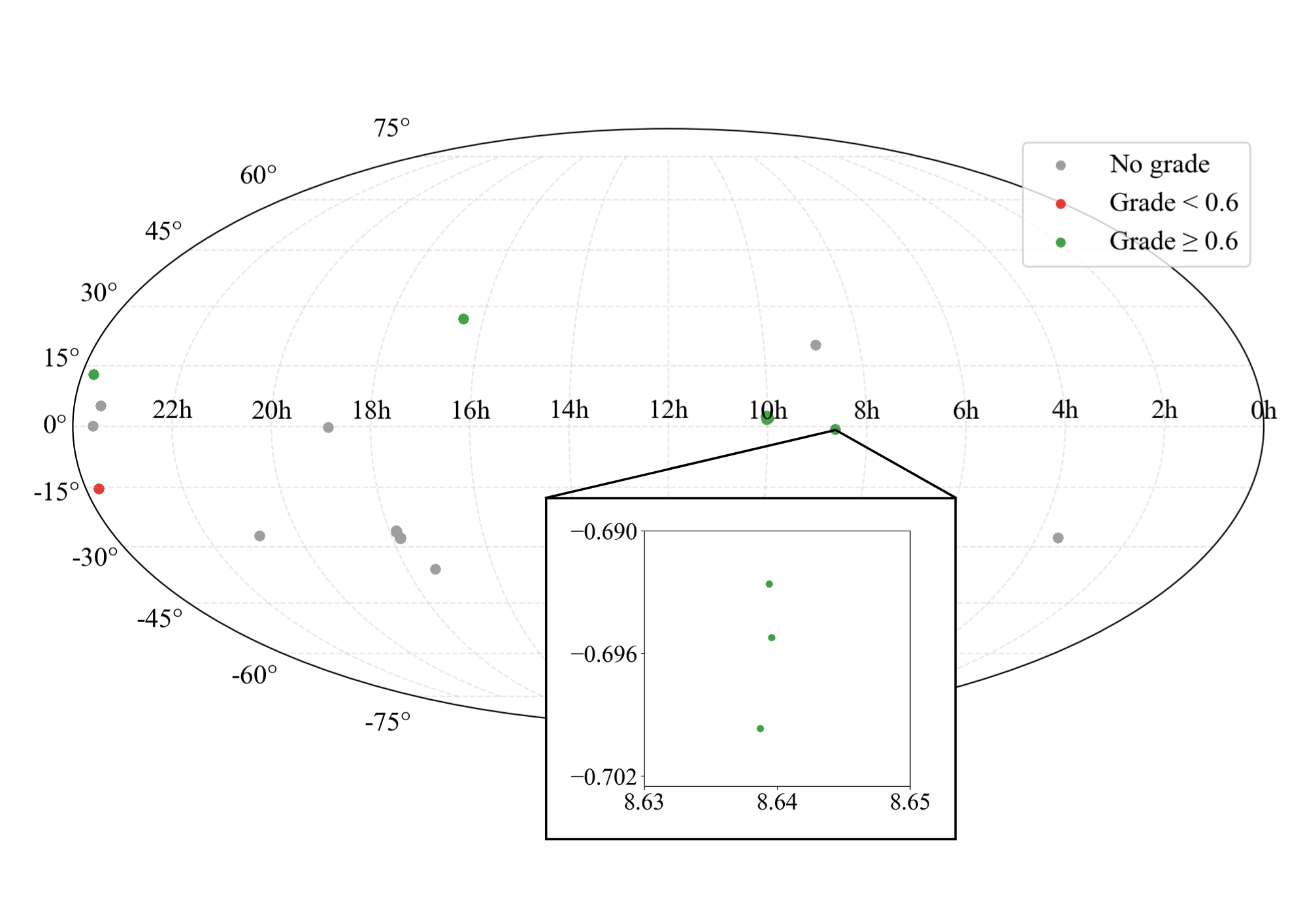}
    \caption{Beamformed pointings across the sky from which the current dataset hits selection was drawn. The beams formed (green, gray and red tick marks - not to scale) show repeated measurements for a small set of regions on the sky. Green ticks show beams searched in observations with $\mathrm{grade}\geq0.6$, red ticks from $\mathrm{grade} < 0.6$ and gray ticks show beams searched in observations with no proceeding calibration observation with which to verify the target observation grade. The zoomed in segment shows individual beams formed on Gaia targets 3073619025268414208, 3073619407521316864 and 3073619338801840000.}
    \label{fig:calibrationgrades}
\end{figure*}

Approximately 116,000 hits occurred when the grade was less than 0.6, 825,000 had a grade greater than or equal to 0.6, and about 1,000,000 had no preceding calibration observation from which a grade could be determined.\footnote{This is not an uncommon occurrence when observations of the target field are short and therefore require only a single calibration at the start of the observing sequence. The COSMIC system is calibrated at this initial phase, but the grade may be low due to changing phase information relative to the previous receiver setup.} Figure~\ref{fig:calibrationgrades} shows the grades and positions of the candidates on the sky.

Given the high-frequency nature of this project, we do not reject any candidates based on their grade, but note that a poor grade will introduce an error in the target location during beamforming. This is expected to reduce the strength of the observed signal relative to its expected Effective Isotropic Radiated Power (EIRP), potentially causing the candidate to be excluded due to insufficient SNR.

\subsection{Source Selection}
As a radio telescope observes at higher frequencies, its field of view (FOV) changes $\propto1/\nu$, such that the higher the frequency the smaller the FOV. Consequently, fewer catalogue sources are visible in each observation (typically only one or two targets per FOV), which limits the number of sources that can be targeted. In this work, we focus on sources from the target selector catalogue, which comprises 31 million stars from the Gaia Data Release II \cite{czech2021breakthrough}. In future studies, however, data obtained through incoherent beams across the full FOV could be compared with a larger dataset to enable a more complete analysis of all sources within the field.

\subsection{The COSMIC Hits Table}
In the real-time pipeline, we search for and record detected hits using the TTDD method implemented in the software \textsc{seticore}. The real-time pipeline using \textsc{seticore} searches for hits at an SNR $>$8$\sigma$ threshold across a Doppler drift range of $\pm$50\,Hz\,s$^{-1}$ \cite{cosmic2023}. Each hit is stored in a table within the COSMIC SQL database, and an up-channelised “postage stamp” - calibrated voltage data from each antenna (hereafter referred to as stamp files) - is saved to disk for further processing and scientific analysis.

Known limitations in the noise estimate used in this search method are discussed in \cite{choza2023breakthrough}, and additional work to evaluate the false-positive rate with improved algorithms in \textsc{seticore}, building on \cite{cosmichighfreq2025AASiposter}, is underway (Tremblay, Myburgh et al., in prep). On average, we find a false-positive rate of roughly 7,000 hits at the 8$\sigma$ threshold per COSMIC observation using \textsc{seticore}. However, across the 696 unique observations in this specific high-frequency dataset, we detected an average of 2,826 hits per observation prior to filtering, with a high degree of variance ranging from 46 to 40,856 hits. This lower hit count relative to the broader COSMIC average is expected, since as previously mentioned, the 25–50 GHz band is largely free of terrestrial radio interference.

The `hits' table in the SQL database contains $30$ fields (see Table \ref{tab:hit_table_descriptions}) as metadata for each hit detected by \textsc{seticore} within the real-time pipeline. These are useful to determine the properties that lead to the hits detection and characterisation, and are used in post processing steps (Section \ref{sec:postprocpipeline}) to separate RFI from potential candidates. One of the signal metadata values is a foreign ID to a ``stamp" entry in the stamp table, allowing us to retrieve the stamp file cutout in which the hit may be re-detected. Stamp files are the calibrated voltage outputs of each antenna that we use in our final layer of filtering, where we analyse whether the hit originated as RFI on a single antenna. In addition, we can use these stamp files as input data to an imaging pipeline (discussed in section \ref{sec:imaging}).

\section{Post-Processing Pipeline}\label{sec:postprocpipeline}
Classification of a hit as non-anthropogenic RFI relies heavily on assumptions about the origin and power of the technosignature source. As already described, a technosignature is assumed to be a very narrow-band ($<$500\,Hz wide \cite{cohen1987narrow}) signal, but it is furthermore assumed to have originated from outside the
solar system. As we expect the transmitter to be on a planet (rotating body) or surrounding a planet and the receiver (our telescope) is also rotating, the change in acceleration from this rotation causes the signal to drift as a function of time and frequency \cite{li2022drift}. However, as many of our targets are toward unknown planetary systems, we complete a blind search using drift rates of $\pm$50\,Hz\,s$^{-1}$, which covers known and unknown phenomena \cite{Li_2023_SETIDR}. 

It also requires noting the known RFI frequency reported by NRAO for K, Ka and Q bands \ref{tab:vla_rfi}\footnote{sourced from \url{https://science.nrao.edu/facilities/vla/docs/manuals/obsguide/rfi}}. These are channels in which we do not search for signals as they are impacted by terrestrial interference.

\begin{table}[h!]
\centering
\caption{Known RFI bands between 25GHz and 50GHz}
\label{tab:vla_rfi}
\begin{tabular}{|c|c|}
\hline
\textbf{Band} & \textbf{Frequencies (GHz)}\\ \hline
Ka  & 29.5--30 \\
Ka & 34.875 \\
Ka & 36.286 \\
\hline
\end{tabular}
\end{table}

\begin{figure*}
    \centering
    \includegraphics[width=0.85\linewidth]{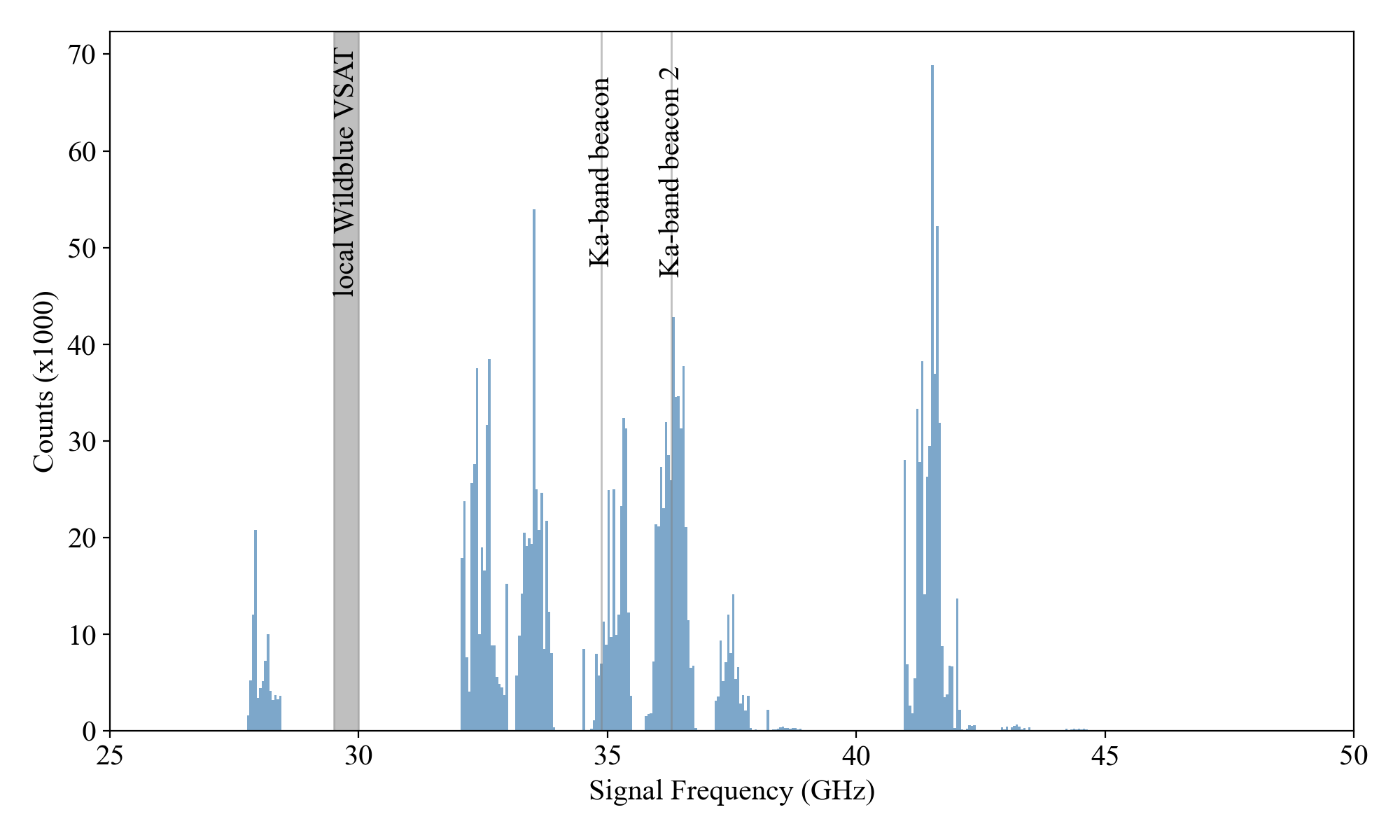}
    \caption{Density histogram showing the distribution of the 1,967,056 detected hits across 25GHz to 50GHz used in this study. Gray bands indicate known continuous RFI regions shown in Table \ref{tab:vla_rfi}. The dominant wide gray band comes from local satellite internet communication via Very Small Aperture Terminal (VSAT) while the two smaller gray bands originate from satellite tracking signals (Ka-band beacons).}
    \label{fig:hitdensityfunction}
\end{figure*}

It was found that none of our candidate signals lie within any known RFI bands and we proceed to subject the candidates to seven bespoke filters described below, building on initial work in filtering high frequency COSMIC signals in \cite{cosmichighfreq2025AASiposter} and (Stiegler, Tremblay in prep). Although these filters were run offline from the real-time pipeline, this paper and  Stiegler, Tremblay in prep, should highlight their performance and motivate for some of the filters being used in the real-time pipeline in the future.

\subsection{Filter One}
We reject signals corresponding to hits within the database that share the same frequency but at different sky positions. This amounts to the rejection of common narrow-band frequency sources, which are most likely RFI, but are not flagged on the NRAO website. This is a common rejection step when multiple beams are formed simultaneously on the sky \cite{vlass2025cosmic}, as this alludes to multiple identical emission sources distributed through space.

\subsection{Filter Two}
We remove hits with a fitted drift rate of exactly zero because stationary narrowband features are overwhelmingly associated with local or instrumental RFI in our observing system. A genuine extraterrestrial signal is expected to exhibit a nonzero apparent drift rate due to the relative acceleration between the transmitter, the Earth, and the telescope. However, we acknowledge that a real signal could appear consistent with zero drift over a short integration if the net line-of-sight acceleration were very small or if the drift were unresolved. This filter therefore reduces the number of false positives at the cost of excluding a small class of possible low-drift signals, and our resulting limits apply only to signals with measurable drift rates $>$0.04\,Hz\,s$^{-1}$.

\subsection{Filter Three}
With only basic geodesic calibration and no noise reduction techniques during the real-time processing, the software running COSMIC produces a high number of false-positive signals (as discussed in a previous section). For very short observations, \textsc{seticore} often detects a hit when there is nothing there (a false positive caused by randomly aligned bright pixels in a dynamic spectra). We impose a SNR requirement for hits observed over fewer time steps (observations with shorter durations than 56 seconds) so that we only keep hits with $\mathrm{timesteps} < 16 = 16\times 0.524288\mathrm{s} \approx 8\mathrm{s}$ and $\mathrm{SNR} > 15$. For a hit with $\mathrm{timesteps} > 16 \approx 8\mathrm{s}$, we require an $\mathrm{SNR} > 10$ to preserve real signals and not just spurious noise. The use of these threshold conditions were arbitrary and used to remove a very large portion of false-positives (see Fig \ref{fig:hitsremovedbyfilter}) whilst also making the research scope of this project tractable. Section \ref{transmitter_limits} details further how a lower SNR bound of $10$ impacts this study.

\subsection{Filter Four}
\begin{figure}
    \centering
    \includegraphics[width=\linewidth]{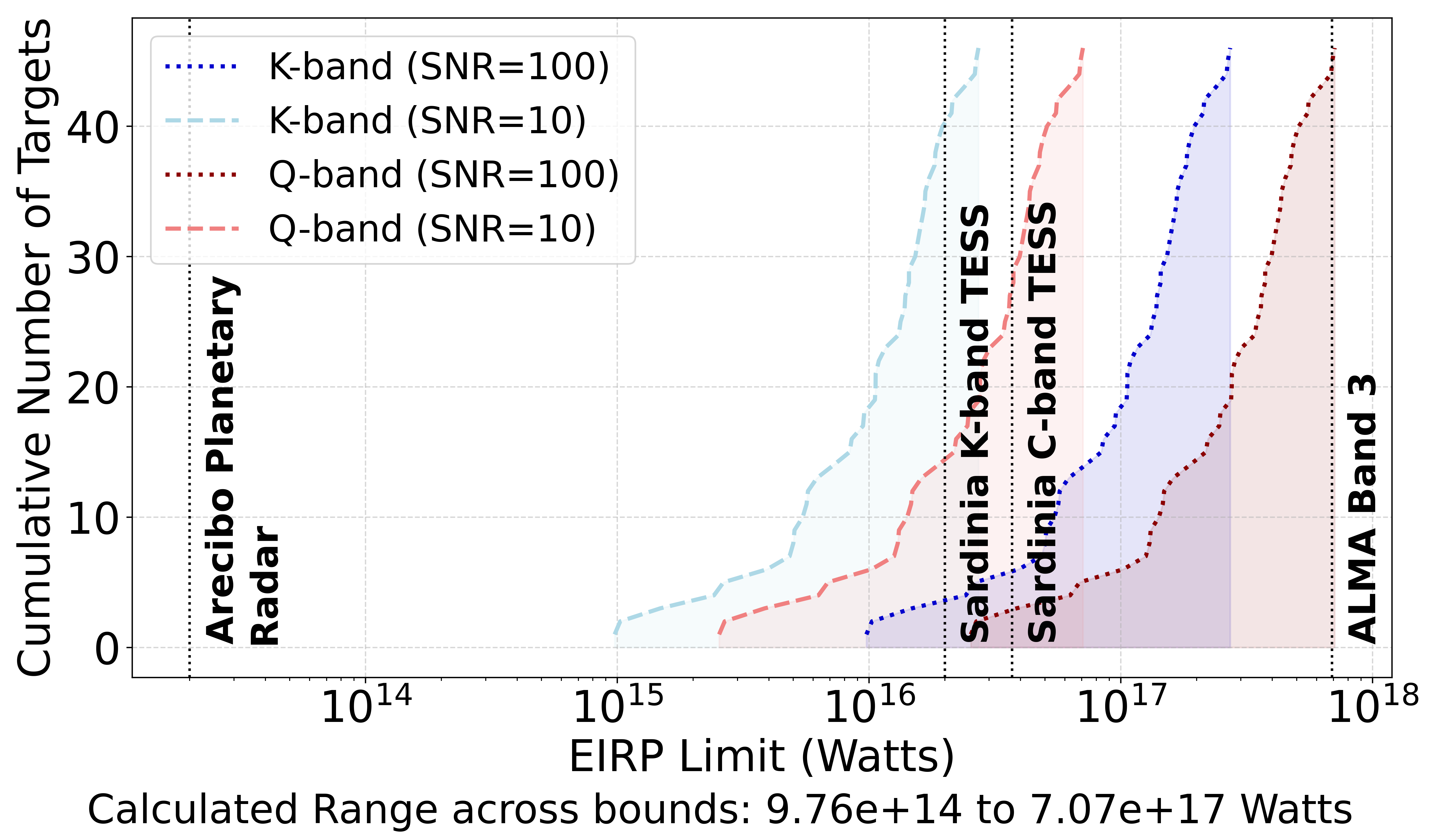}
    \caption{Plot showing the cumulative distribution function for each individual target from Table \ref{tab:eirp_fixed} across the lower and and upper frequency bands (K and Q) and lower and upper SNR thresholds (10 and 100) for our study. Dashed vertical lines indicate the $\mathrm{EIRP}_{\min}$ thresholds from high frequency technosignature search studies \cite{Sardinia}, \cite{Siemion2013} and \cite{mason2025conducting}. Source QSO J0854+2006 from table \ref{tab:eirp_fixed} is not included in this plot as its outlier distance distorts this plot significantly.}
    \label{fig:eirpcdf}
\end{figure}

In addition, signals originating from interstellar distances (i.e., beyond the Solar System) are attenuated in total intensity according to the inverse-square law. We set an upper SNR threshold cutoff of $100$ for this filter to ensure the high fidelity of our candidate sample and to aggressively mitigate false positives from anthropogenic RFI and instrumental artifacts. As will be shown in Section \ref{transmitter_limits}, Figure \ref{fig:eirpcdf}, this threshold necessitates an exceptionally high $EIRP_{max}$. When normalized to Arecibo Power Units ($L_A$), our search requires a transmitter roughly 500 to 35,000 times more luminous than the Arecibo planetary radar power. While this establishes a strict energy requirement across the K, Ka and Q band targets, it guarantees that surviving candidates represent highly deliberate, civilization-scale (Kardashev Type I) engineered emissions that are robust against statistical fluctuations and local noise. This filter will be refined in future work and perhaps expanded to include the possibility of more advanced civilisations, but will require that the other filters used are better honed to reject emission of anthropogenic origin.

\subsection{Filter Five}
If the Doppler motion is tracked between two rotating bodies, the resulting Doppler shift varies across observations in a predictable sinusoidal pattern. However, when a signal undergoes both coarse and fine channelization, we expect some drift across several channels within the detection window. To identify frequency modulation, we assume that all hits within 10 Hz of a given hit in a single observation are modulations of that signal. We then remove these hits and their suspected modulations, retaining only ``lonely" hits with no other detections within 10 Hz. The choice of five frequency bins (each fine-channel being 2 Hz) was likely an overly safe precaution, but it ensured that we allowed far fewer false-positives through and reduced the problem set significantly.

\subsection{Filter Six}
We apply a continuity filter in which, for each hit, we examine subsequent observations of the same position to check whether the signal persists along the expected frequency trajectory determined by the earlier drift rate. Hits that do not meet this criterion are discarded as isolated events. Although there remains a slight possibility that a genuine source may be observed only once-leading to some missed candidates-this approach substantially increases the reliability of the hits that pass the filter.

\subsection{Filter Seven}\label{sec:incoherent_filter}
Following what was done in Section 3.9 of \cite{cosmic2023} we apply a filter to our coherent hits such that if

\begin{equation}
    \mathrm{coherent\_S/N} \leq \sqrt{N_{antennas}} \times \mathrm{incoherent\_S/N}
\end{equation}

then the signal is most likely RFI. Both incoherent ($\mathrm{incoherent\_S/N}$) and coherent ($\mathrm{coherent\_S/N}$) signal power are calculated by \textsc{seticore} and stored for every entry of `hits' in the database.

\section{Results}
Between 01 February 2024 and 30 April 2025, a total of 1,967,056 hits between 25\,GHz and 50 \, GHz were recorded. For this experiment, only hits located within the coherent beams were studied. Although techniques for filtering analysis of the incoherent beam hits would not be too dissimilar from those described in this paper, this choice suitably narrowed the focus of this experiment.

Approximately 600,000 of the recorded hits were from coherent beams formed towards 45 $Gaia$ targets. These targets were observed for a total of $\approx59.25$ hours during this period. The remainder of the hits were from a coherent beam formed at the phase centre, which occurs when the field has no known sources. In this scenario, we note the right ascension, declination, array configuration and expected synthesized beamwidth at the time of the observation (Table \ref{tab:synthesized_beam}) and do a $SIMBAD$\footnote{\url{https://simbad.u-strasbg.fr/simbad/}} cone-search. For this dataset, we find that 6 sources:
\begin{itemize}
    \item \,[VV2006\,] J095858.7+020138
    \item COSMOS2015 873756
    \item COSMOS2015 818760
    \item V* R Aqr
    \item QSO J0854+2006
    \item ACS-GC 90043285
\end{itemize}
were co-located with the coherent beam formed at phase centre, and were collectively observed for $\approx 116.5$ hours.

As shown in Table \ref{tab:vla_rfi} and Figure \ref{fig:hitdensityfunction}, only a small fraction of the band was impacted by known persistent RFI. However, there are frequencies where there is a higher density of hits compared to the other frequencies. 

After the hits were analysed against the criteria for each of the filters described in Section \ref{sec:postprocpipeline}, no hits remained. The pipeline did not detect technosignatures.

With no remaining hits, this suggests either an incorrect assessment of what a technosignature looks like or the frequency range has more RFI than suggested in simple ``RFI" scans provided by NRAO. At this time, none of these steps are clear indicators of either scenario, so are not adopted into the real-time pipeline. 

\subsection{Transmitter Limits}\label{transmitter_limits}
To determine how bright a transmitted signal would need to be from each source in order for us to detect it, we compute the minimum equivalent isotropic power (EIRP$_\mathrm{{min}}$) for each source,

\begin{equation}
    EIRP_\mathrm{{min}} = 4 \pi d^2 F_\mathrm{{min}}
\end{equation}

where the relationship to the detectable power is based on a distance squared ($d$). The value of $F_\mathrm{{min}}$ is determined by dividing the minimum flux density by the transmission signal bandwidth. This changes the equation into the format: 

\begin{equation}
\mathrm{EIRP}_{\min}\,[\mathrm{W}] \;\approx\; 
1.20 \times 10^{8} \;
\left( \frac{d}{\mathrm{pc}} \right)^{2} \;
\left( \frac{S_{\min}}{\mathrm{Jy}} \right) \;
\left( \frac{\Delta\nu}{\mathrm{Hz}} \right)
\end{equation}

Since technosignatures are searched for in beamformed data, we use the following equation to calculate the lower flux density limit ($S_{limit}$):
\begin{equation}\label{eq:slimit}
S_{limit} = \frac{\mathrm{SEFD}}{B_e\sqrt{n_{pol}\times n \times t_{int} \times \Delta\nu}}\times SNR
\end{equation}
where $Be=0.9$ is the beam efficiency of COSMIC\cite{cosmic2023}, $n_{pol} = 2$ is the number of polarisations, $n=25$ is the number of operating antenna, $t_{int}=56$s is the integration time, $\Delta\nu=2$Hz is the channel width, and $\mathrm{SEFD}$ is the system equivalent flux density found to equal $500$Jy at K-band and $1300$Jy and Q-band\footnote{\url{https://science.nrao.edu/facilities/vla/docs/manuals/oss/performance/sensitivity}}.

For this experiment, using equation \ref{eq:slimit}, we calculate the minimum flux density limits at 10$\sigma$ (see filter three) to be 74.24\,Jy\,beam$^{-1}$ at 25\,GHz and 193.02\,Jy\,beam$^{-1}$ at 50\,GHz.

\begin{figure}
    \centering
    \includegraphics[width=\linewidth]{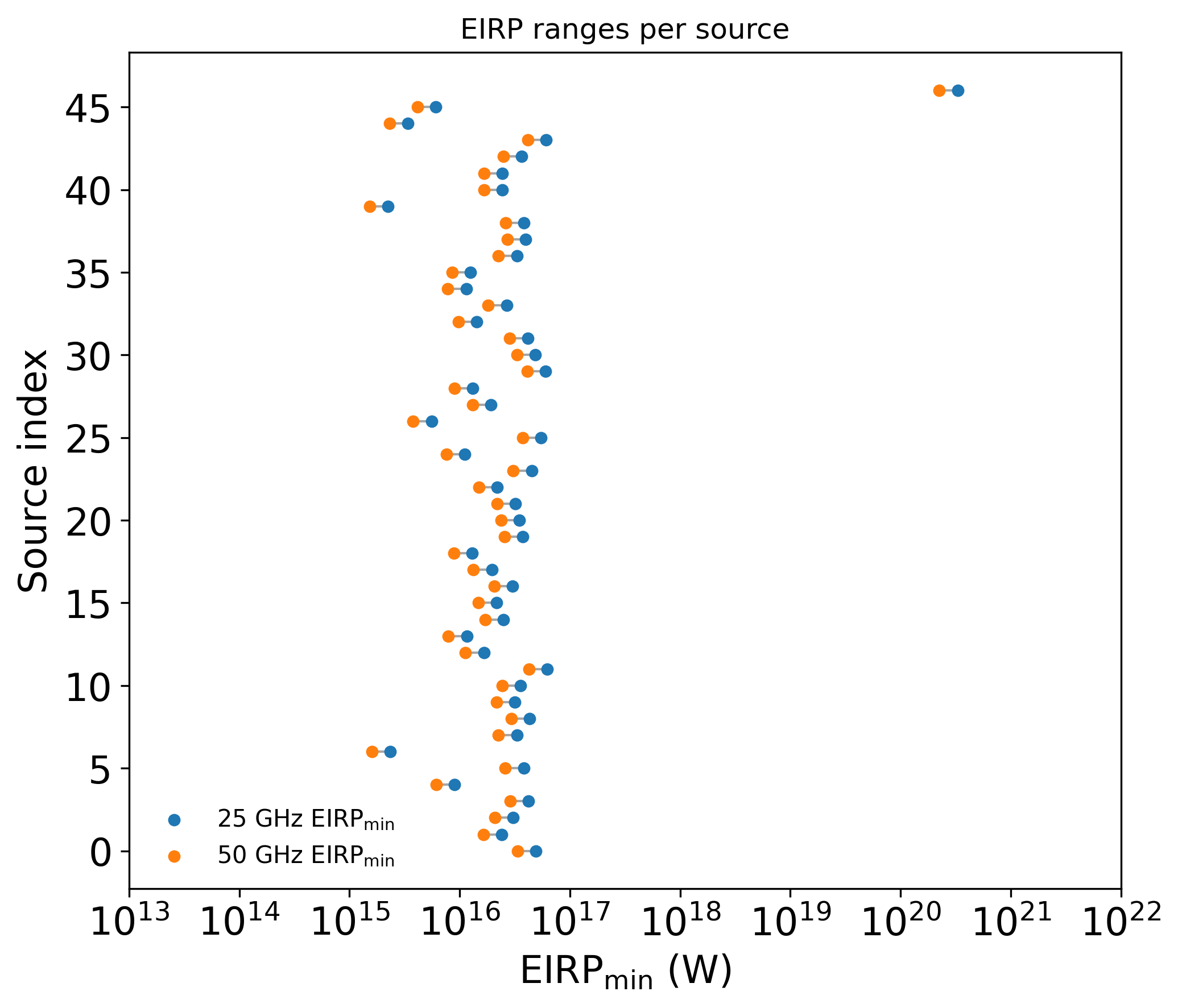}
    \caption{Plot showing the distribution of EIRP limit values for a transmitter sitting within each stellar system.}
    \label{fig:EIRP}
\end{figure}

As shown in Figure \ref{fig:EIRP} and Table \ref{tab:eirp_fixed}, the EIRP$_\mathrm{{min}}$ values range from 2.24$\times$10$^{15}$ to 3.31$\times$10$^{20}$\,W. Based on the distances calculated using parallax in \cite{czech2021breakthrough} for the $Gaia$ sources, the sources range in distance between 234 and 90090\,pc. Using the power value of the Arecibo Planetary Radar (L$_A$) of $\approx2\times$10$^{13}$\,W from Siemion et al. \cite{Siemion2013}, we can calculate the ratio (EIRP$_\mathrm{{min}} \div$L$_{A}$) of this value for each source in our survey. As a result, we established a value range between 112 and 1.655$\times$10$^{7}$.

Accounting for filters 3 and 4 \ref{sec:postprocpipeline}, this study has imposed SNR rejection thresholds of $\leq 10$ and $\geq100$. This in turn translates to an $\mathrm{EIRP}_{\min}$ rejection limit of $<9.76\times10^{14}\mathrm{W}$ and $\mathrm{EIRP}_{\min}$ rejection limit of $>7.06\times10^{17}\mathrm{W}$.

\subsection{Visual Inspection of Candidates}
\begin{figure}
    \centering
    \includegraphics[width=\linewidth]{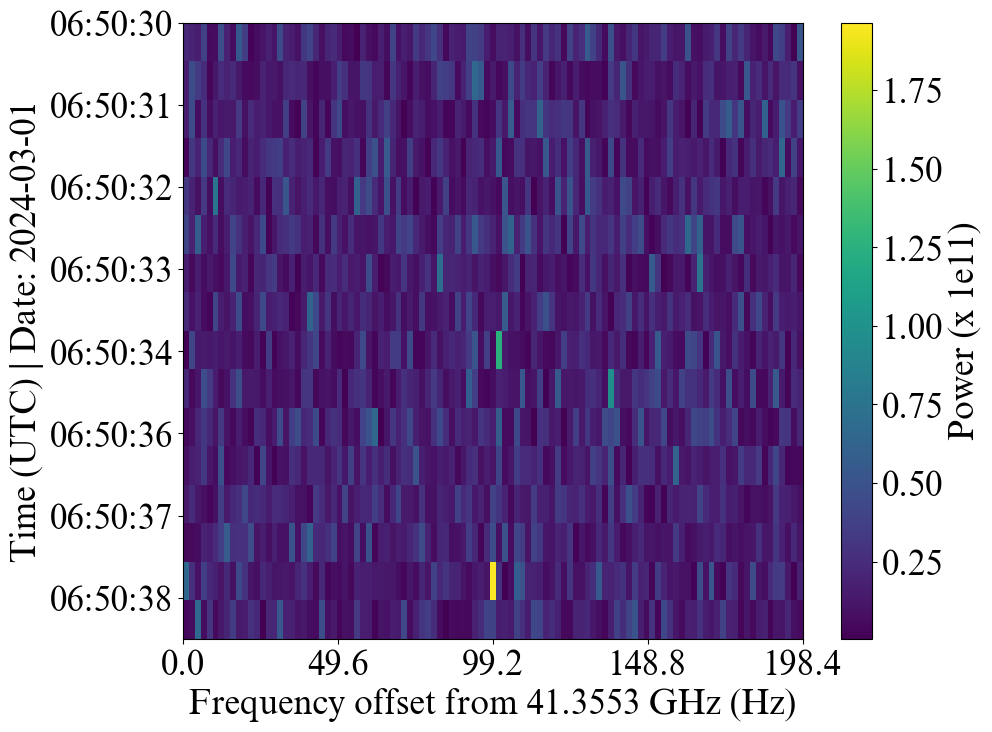}
    \caption{Drift rate -0.727595761 signal that passed through all six filters before being eliminated by filter 7 for having a measured coherent power not sufficiently more powerful than the measured incoherent power (see Section \ref{sec:incoherent_filter}). Power units are in $\mathrm{ADC}^2$ i.e. $|voltage|^2$ as calculated by \textsc{seticore}}
    \label{fig:waterfall1}
\end{figure}

For all hits that remain after the filters, the data are visualised for inspection of the signal characteristics. However, since no hits passed through the filters, we examined a hit that passed through all but one of the filters in Figure \ref{fig:waterfall1} as an example of this step. As noted in the figure caption, this hit was algorithmically eliminated by Filter 7 because its coherent power was not sufficiently greater than its incoherent power, marking it as non-localized interference. The identification of the hit by the \textsc{seticore} software is due to the step where the signal is time averaged after de-drifting. The bright pixel shown in the dynamic spectra (waterfall plot) connected with other singular bright pixels in other time bins, creates a signal that is interpreted to be a candidate by the software. However, this does not demonstrate what we expect a technosignature to look like where all time bins would have a signal with a linearly changing frequency. 

\begin{figure}
    \centering
    \includegraphics[width=\linewidth]{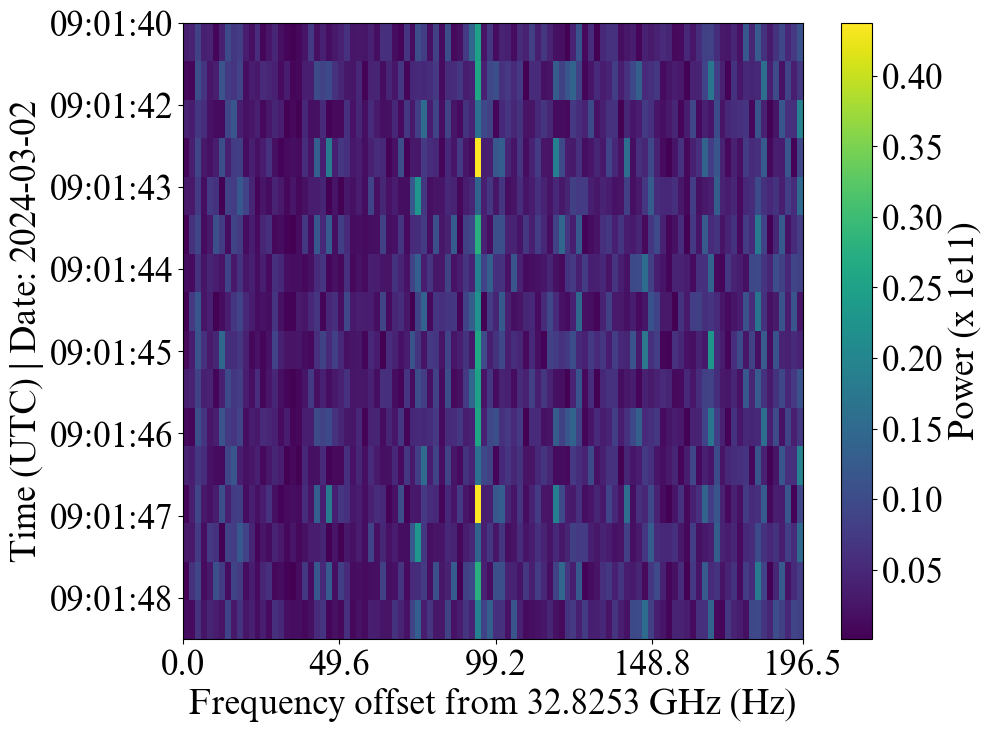}
    \caption{Drift rate 0 signal passed through filter 7, but not through filter 2 due to having a drift rate precisely equal to zero. Power units are in $\mathrm{ADC}^2$ i.e. $|\mathrm{voltage}|^2$ as calculated by \textsc{seticore}}
    \label{fig:waterfall2}
\end{figure}

Another example is shown in Figure \ref{fig:waterfall2}, where we demonstrate one of the signals that passed through all of the filters except it has a drift rate of zero. We would expect that a technosignature coming from another planetary system would have a signal drifting in time and frequency due to Doppler acceleration. Therefore, the pipeline correctly removed this signal. However, it does demonstrate a signal that is persistent in time, although with a changing brightness, as shown in the changes in colour along the signal path.

\subsection{Imaging \& Correlation}\label{sec:imaging}
Another test to determine whether a signal is of astronomical origin would be to image the field. We would expect a technosignature to appear as a point source in an interferometric image toward the position in which we beamformed. In contrast, RFI would appear as structured interference not localised to a single position on the sky.

\subsubsection{Correlation}
To image the saved data after the real-time pipeline, we correlate the calibrated voltage data in each stamp file using the Breakthrough Listen Interferometry Package (BLRI) \footnote{\url{https://github.com/MydonSolutions/BLRI}} and then convert the resultant $UVH5$ file to a CASA Measurement Set ($MS$) \footnote{\url{https://casaguides.nrao.edu/index.php?title=Measurement\_Set\_Contents}} using \textsc{pyUVData}\footnote{\url{https://pyuvdata.readthedocs.io/en/latest/}}.

Correlations are computed for each unique pair of antennas for which there is data in the stamp file, producing `baselines'. These baselines are written into a $UVH5$ formatted file along with a $UVW$-array calculated for the observation. Included are further metadata on the telescope information and details on antenna positions and phase centre \footnote{\url{https://github.com/RadioAstronomySoftwareGroup/pyuvdata/blob/main/docs/references/uvh5\_memo.pdf}}.

\begin{figure}
    \centering
    \includegraphics[width=\linewidth]{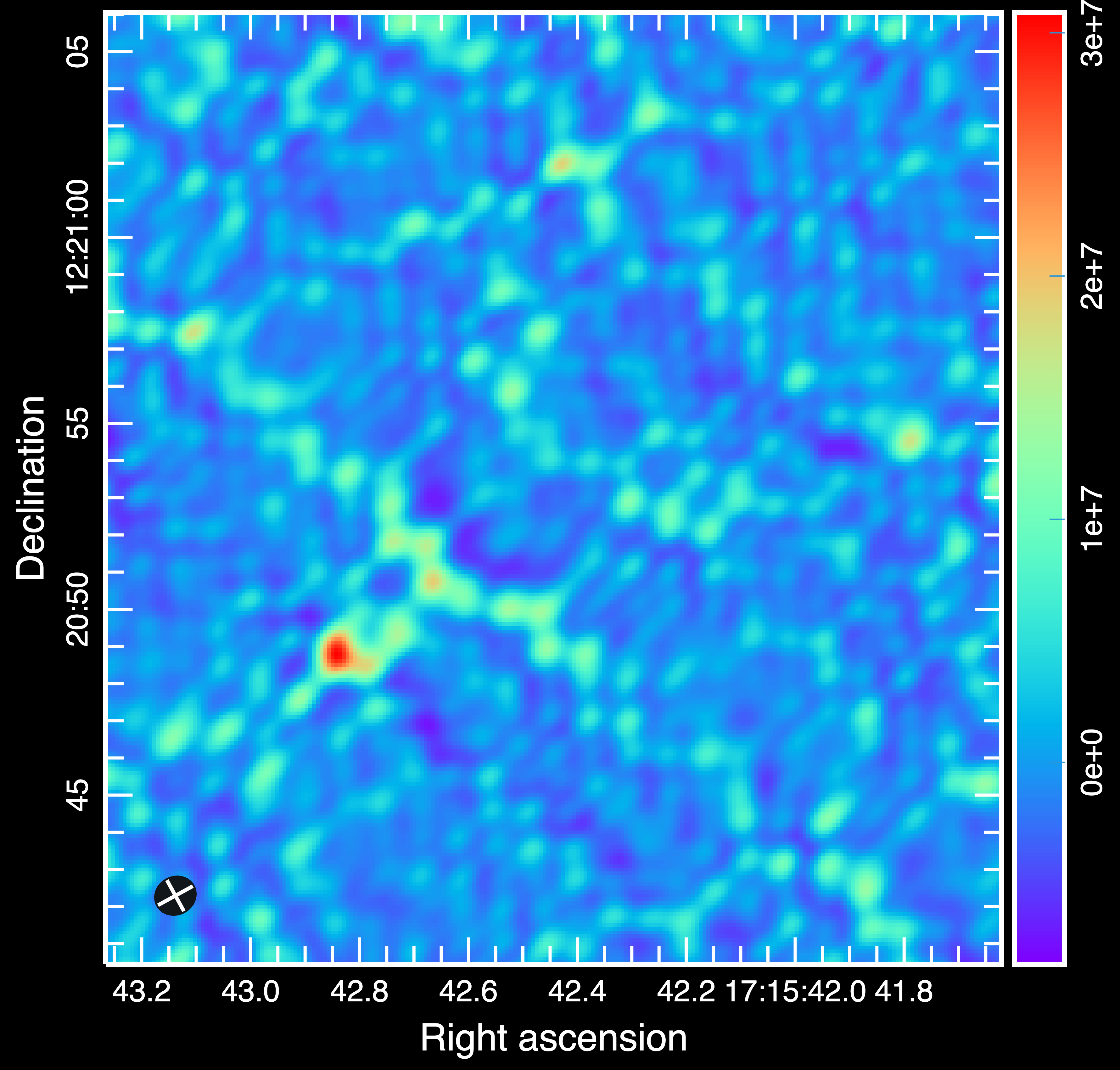}
    \caption{A preliminary image towards Voyager 1 created using visibilities correlated from COSMIC raw voltages. The synthesised beam (resolving element) of the VLA for this observation is shown in the bottom left of the image, with the white cross indicating the major and minor axis of the elliptical Gaussian. This image was created from five $2\,\rm{Hz}$ channels at a central frequency of $8420.4191\,\rm{MHz}$. A candidate for voyager can be seen towards the bottom left as a red excess close to the shape of the synthesised beam. Further validation of this detection is ongoing. Only the left circular polarisation was imaged.}
    \label{fig:image}
\end{figure}

\subsubsection{Imaging}
For each candidate signal, the postage-stamp voltage data is alternatively translated into correlated visibilities across a narrow frequency window surrounding the detection. Most interferometric imaging software operates on \textsc{CASA} Measurement Set (MS) data products, and therefore we use the \texttt{pyuvdata} package to convert the native \texttt{UVH5} format into an MS file. We then image individual spectral channels using the CASA task \texttt{tclean}, integrating over the duration of the observation. Prior to imaging, the visibilities may be de-drifted using the Doppler drift rate measured by \texttt{seticore}, concentrating the signal power into fewer frequency channels and thereby increasing the image signal-to-noise ratio.

As a demonstration of this technique, we observed Voyager~1 with the VLA at X-band on 2023/04/26 . At the time of observation, Voyager~1 was calculated to be located at J2000 coordinates RA = 17$^{\rm h}$ 15$^{\rm m}$ 42$^{\rm s}$, Dec = $+$12$^\circ$ 20$'$ 53.30$''$ as obtained from the JPL Horizons ephemeris through Astroquery. The spacecraft was detected at RA = 17$^{\rm h}$ 15$^{\rm m}$ 42.8$^{\rm s}$, Dec = $+$12$^\circ$ 20$'$ 49$''$. The spacecraft's narrowband telemetry signal provides a useful test case for a sky-localized artificial transmitter. As shown in Figure~\ref{fig:image}, the emission from Voyager~1 is recovered as an unresolved point source whose morphology is consistent with the synthesized beam of the array. In contrast, terrestrial and instrumental radio-frequency interference typically produces spatially extended or structured emission that is not localized to a single position on the sky, making interferometric imaging a powerful discriminator during candidate verification.

\section{Discussion \& Future Work}
\begin{figure}
    \centering
    \includegraphics[width=\linewidth]{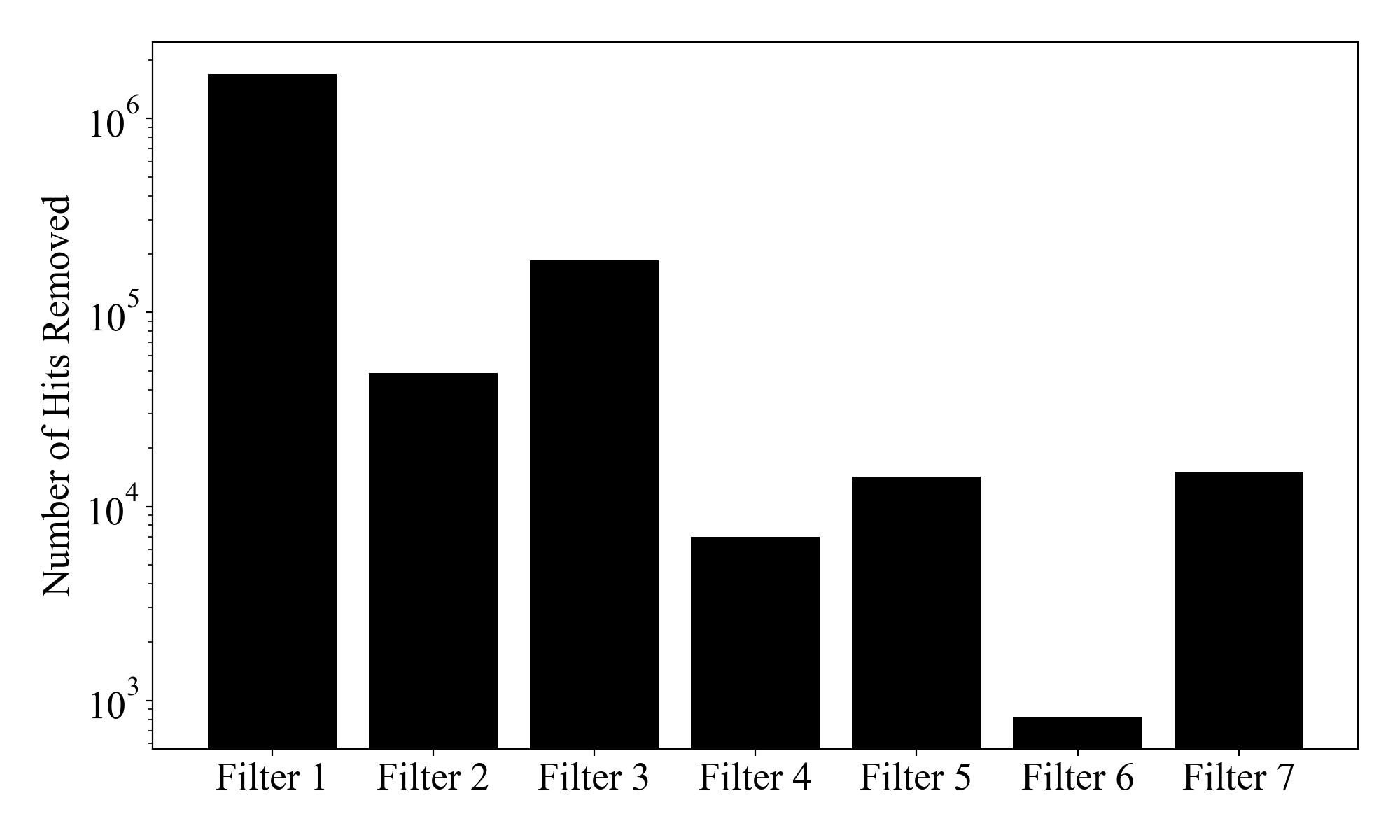}
    \caption{Number of candidate hits eliminated by each filtering stage.}
    \label{fig:hitsremovedbyfilter}
\end{figure}

Figure \ref{fig:hitsremovedbyfilter} shows that over 85\% of the selected hits dataset had multiple hits corresponding to the same frequency at differing positions in the sky and were removed by filter 1. This is a strong indication for RFI and speaks to the ever increasing sources of RFI even in well protected zones such as the VLA. We also note that after passing through all filters preceding filter six, filter six removes only 4\% of the remaining candidates. Filter six aims to detect narrow-band transient candidates most likely to arise from transient RFI or systematic errors, but, because this filter is applied sixth in the filter queue, candidates entering this filter must be drifting and of reasonable power levels. We expect this class of candidates to be rare and as such should look to refine filter six considering its impact.

Although no new signals were determined, this paper represents the first published results for a search between 25--50\,GHz. As we do not know what frequency a technosignature would be transmitted at, covering as much of the electromagnetic spectrum is important. This search also represents a high time (500\,ms) and frequency resolution (2\,Hz) search to identify narrow-band drifting signals. Due to the smaller wavelengths involved in higher-frequency receivers, their design accuracy must be greater, making them more expensive from an economic standpoint; making this a rare opportunity in SETI.

The results of this survey place limits on high-frequency transmitters comparable to those reported in the 18 GHz survey conducted with the Sardinia Radio Telescope \cite{Sardinia}. That study focused on $TESS$ targets and several observations of the Galactic Centre, whereas our survey primarily targets sources from the $Gaia$ catalogue and includes stars that may not have known planetary systems. Petigura et al. \cite{Petigura_2013} estimate that up to 20\% of Sun-like stars host Earth-sized planets, indicating that observations of low-mass $Gaia$ stars add significant value to SETI.

Looking ahead, we plan to integrate this pipeline into a more efficient software package for periodic post-processing of data collected by the real-time system. We are also preparing a paper presenting results from a more extensive high-frequency study, including a detailed analysis of interference detected in this regime and a full assessment of the pipeline’s false-positive rate.

Finally, we aim to improve the imaging capabilities of the system, which will provide additional constraints on potential candidate signals across all frequencies observed with the VLA. 

\section{Conclusions}
In this study, we observed 45 stars from the Gaia catalogue with the VLA using the COSMIC digital back end to search for candidate signals. To distinguish radio-frequency interference (RFI) from signals of astronomical origin, we developed a pipeline consisting of a series of logical filters. Although no signals satisfied all of these criteria, we also built a visualization pipeline to support analysis of future data from the ongoing COSMIC experiment.

Overall, this work demonstrates the capability to search for technosignatures across a broader frequency range than has traditionally been explored. It also showcases a rapid post-processing pipeline that enables candidate searches to be carried out quickly and efficiently -- an essential feature given that the system ingests terabytes of data each week.

\section*{Acknowledgements}

We gratefully acknowledge the foundational support from
John and Carol Giannandrea that has made COSMIC possible.
We acknowledge additional support from other donors,
including the Breakthrough Prize Foundation under the
auspices of Breakthrough Listen. The National Radio Astronomy Observatory is a facility of the National Science
Foundation operated under a cooperative agreement with
Associated Universities, Inc.  

\newpage
\onecolumn
\section*{Appendix}

\vspace{5\baselineskip}

\begin{center}
    \begin{tabular}{llll}
    \hline
         Start date & End date & Configuration & Synthesized Beamwidth \\
         UTC & UTC & & arcsec\\
    \hline
    2024-01-16 10:29:00 & 2024-01-21 17:40:00 & D & 2.1\\
    2024-01-21 17:40:00 & 2024-04-22 09:07:00 & C & 0.63\\
    2024-04-26 13:47:00 & 2024-09-16 08:51:59 & B & 0.19\\
    2024-09-20 11:48:27 & 2024-10-07 14:12:37 & BnA & NaN\\
    2024-10-14 15:55:31 & 2025-02-03 09:22:11 & A & 0.059\\
    2025-02-26 17:01:17 & 2025-06-01 00:00:00 & D & 2.1\\
    \hline
    \end{tabular}
    \captionof{table}{Array configuration and Ka synthesized beamwidth between 2024-01-16 and 2025-06-01. Synthesized beamwidths are drawn from NRAO resource: \url{https://science.nrao.edu/facilities/vla/docs/manuals/oss/performance/resolution}}
    \label{tab:synthesized_beam}
\end{center}

\twocolumn


\begin{table*}[ht]
\centering
\resizebox{\textwidth}{!}{%
\begin{tabular}{lrrrlll}
\hline
Source ID & RA & Dec & Distance& EIRP$_{\mathrm{min}}$ & EIRP$_{\mathrm{min}}$ & No. of Obs. \\
& (hours) & (degrees) & (pc) & at 25\,GHz (W) & at 50\,GHz (W) &\\
\hline
Gaia DR3 1304986578365581440 & 246.454 & 26.831 & 1098.99 & 4.92e+16 & 3.35e+16 & 47\\
Gaia DR3 4057717120813215872 & 267.390 & -27.735 & 770.65 & 2.42e+16 & 1.65e+16 & 1\\
Gaia DR3 4057670288487780096 & 267.481 & -27.751 & 866.57 & 3.06e+16 & 2.09e+16 & 1\\
Gaia DR3 4057670249822469376 & 267.448 & -27.753 & 1016.55 & 4.21e+16 & 2.87e+16 & 1\\
Gaia DR3 4057716983374266240 & 267.363 & -27.750 & 470.77 & 9.03e+15 & 6.15e+15 & 1\\
Gaia DR3 4057717155172970880 & 267.344 & -27.747 & 967.68 & 3.82e+16 & 2.60e+16 & 1\\
Gaia DR3 4057711348377192704 & 267.328 & -27.751 & 240.42 & 2.36e+15 & 1.60e+15 & 1\\
Gaia DR3 4057711314017867648 & 267.306 & -27.755 & 900.65 & 3.31e+16 & 2.25e+16 & 1\\
Gaia DR3 4057670185408565888 & 267.458 & -27.770 & 1030.32 & 4.33e+16 & 2.95e+16 & 1\\
Gaia DR3 4057669360774821760 & 267.485 & -27.785 & 882.82 & 3.18e+16 & 2.16e+16 & 1\\
Gaia DR3 4057669360774820224 & 267.488 & -27.784 & 936.39 & 3.57e+16 & 2.43e+16 & 1\\
Gaia DR3 4057669738731973248 & 267.429 & -27.785 & 1237.58 & 6.24e+16 & 4.25e+16 & 1\\
Gaia DR3 4057717872419963136 & 267.428 & -27.679 & 638.19 & 1.66e+16 & 1.13e+16 & 1\\
Gaia DR3 4057717670569033344 & 267.435 & -27.690 & 533.60 & 1.16e+16 & 7.91e+15 & 1\\
Gaia DR3 4057718559614735616 & 267.370 & -27.676 & 784.70 & 2.51e+16 & 1.71e+16 & 1\\
Gaia DR3 4057717597541561216 & 267.438 & -27.701 & 729.85 & 2.17e+16 & 1.48e+16 & 1\\
Gaia DR3 4057717734980975616 & 267.399 & -27.704 & 859.62 & 3.01e+16 & 2.05e+16 & 1\\
Gaia DR3 4057671869035760384 & 267.486 & -27.724 & 693.08 & 1.96e+16 & 1.33e+16 & 1\\
Gaia DR3 4057672075194206208 & 267.474 & -27.714 & 563.82 & 1.30e+16 & 8.83e+15 & 1\\
Gaia DR3 4057670593419907968 & 267.458 & -27.718 & 957.51 & 3.74e+16 & 2.55e+16 & 1\\
Gaia DR3 4057718014166454016 & 267.327 & -27.722 & 926.77 & 3.50e+16 & 2.38e+16 & 1\\
Gaia DR3 4057670387261449600 & 267.471 & -27.739 & 885.78 & 3.20e+16 & 2.18e+16 & 1\\
Gaia DR3 4057668948457957376 & 267.463 & -27.808 & 733.88 & 2.19e+16 & 1.50e+16 & 1\\
Gaia DR3 4057669738731963136 & 267.429 & -27.800 & 1050.24 & 4.49e+16 & 3.06e+16 & 1\\
Gaia DR3 4057668982817709312 & 267.442 & -27.805 & 521.74 & 1.11e+16 & 7.56e+15 & 1\\
Gaia DR3 4057669665704274176 & 267.400 & -27.806 & 1159.33 & 5.48e+16 & 3.73e+16 & 1\\
Gaia DR3 4057664000655691520 & 267.369 & -27.800 & 369.15 & 5.55e+15 & 3.78e+15 & 1\\
Gaia DR3 4057710863030591744 & 267.331 & -27.799 & 686.70 & 1.92e+16 & 1.31e+16 & 1\\
Gaia DR3 4057710897390306688 & 267.313 & -27.806 & 567.47 & 1.31e+16 & 8.94e+15 & 1\\
Gaia DR3 4057669051537149056 & 267.498 & -27.817 & 1217.26 & 6.04e+16 & 4.11e+16 & 1\\
Gaia DR3 4057663760137861120 & 267.376 & -27.822 & 1094.06 & 4.88e+16 & 3.32e+16 & 1\\
Gaia DR3 4057663893268143360 & 267.324 & -27.824 & 1012.71 & 4.18e+16 & 2.85e+16 & 1\\
Gaia DR3 4057663828857338368 & 267.349 & -27.838 & 591.24 & 1.42e+16 & 9.71e+15 & 1\\
Gaia DR3 4057663515311002368 & 267.316 & -27.831 & 808.90 & 2.67e+16 & 1.82e+16 & 1\\
Gaia DR3 4057662896835646592 & 267.422 & -27.844 & 530.70 & 1.15e+16 & 7.82e+15 & 1\\
Gaia DR3 4057662694985555968 & 267.438 & -27.859 & 553.96 & 1.25e+16 & 8.52e+15 & 1\\
Gaia DR3 4057663588339157376 & 267.379 & -27.844 & 900.56 & 3.30e+16 & 2.25e+16 & 1\\
Gaia DR3 4057662832424908160 & 267.384 & -27.849 & 987.15 & 3.97e+16 & 2.71e+16 & 1\\
Gaia DR3 4057662832424901760 & 267.391 & -27.858 & 970.84 & 3.84e+16 & 2.62e+16 & 1\\
Gaia DR3 3073619025268414208 & 129.581 & -0.700 & 234.40 & 2.24e+15 & 1.53e+15 & 1\\
Gaia DR3 3073619407521316864 & 129.591 & -0.693 & 772.94 & 2.43e+16 & 1.66e+16 & 1\\
Gaia DR3 3073619338801840000 & 129.593 & -0.695 & 773.40 & 2.44e+16 & 1.66e+16 & 1\\
Gaia DR3 4066907835429532416 & 268.037 & -26.206 & 946.38 & 3.65e+16 & 2.49e+16 & 13\\
Gaia DR3 4066997716194441344 & 267.935 & -25.849 & 1223.79 & 6.10e+16 & 4.16e+16 & 1\\
Gaia DR3 4266157484359852288 & 282.847 & -0.198 & 288.71 & 3.40e+15 & 2.31e+15 & 1\\
V* R Aqr & 355.956 & -15.285 & 385.64 & 6.06e+15 & 4.13e+15 & 4\\
QSO J0854+2006 & 133.704 & 20.109 & 90090.09 & 3.31e+20 & 2.25e+20 & 1\\
\hline
\end{tabular}
}
\caption{EIRP values with fixed calculations.}
\label{tab:eirp_fixed}
\end{table*}

\begin{table*}[htbp]
\centering
\begin{tabularx}{\textwidth}{lX}
\toprule

\textbf{Field} & \textbf{Description} \\
\midrule
\texttt{id} & Unique numeric identifier for the hit entity. \\
\texttt{beam\_id} & Foreign identifier of the beam entity in which the signal was detected (necessary to re-detect the hit from the stamp data which is not beamformed). \\
\texttt{observation\_id} & Foreign identifier of the observation entity. \\
\texttt{stamp\_id} & Foreign identifier of the stamp entity. \\
\texttt{tuning} & Receiver tuning (AC or BD in 8-bit mode). \\
\texttt{subband\_offset} & Frequency offset of the subband relative to the channel center. \\
\texttt{file\_uri} & Uniform Resource Identifier listing the location of the hits file. \\
\texttt{file\_local\_enumeration} & Local file sequence index. \\
\texttt{signal\_frequency} & Calculated topocentric frequency of the detected signal (MHz). \\
\texttt{signal\_index} & The fine-frequency channel index that the hit start in, within the coarse-frequency channel. \\
\texttt{signal\_drift\_steps} & The drift rate expressed as an integer number of channel steps over the full observation time. \\
\texttt{signal\_drift\_rate} & The measured frequency drift rate of the signal (Hz/s). \\
\texttt{signal\_snr} & Signal-to-Noise Ratio (SNR) of the detected candidate. \\
\texttt{signal\_coarse\_channel} & The specific coarse channel index within which the narrow-band signal was detected. \\
\texttt{signal\_beam} & The beam number in which the signal was recorded (index to which beam in the beam entry). \\
\texttt{signal\_num\_timesteps} & The duration of the signal expressed as the number of time integrations it spans. \\
\texttt{signal\_power} & Integrated coherent power of the detected signal. \\
\texttt{signal\_incoherent\_power} & Integrated incoherent power of the signal (signal power measured with no beamforming). \\
\texttt{source\_name} & Name of source or ``PHASE\_CENTER'' if beam not directed at a astronomical target. \\
\texttt{fch1\_mhz} & Center frequency of the first channel in the data file (MHz). \\
\texttt{foff\_mhz} & Channel bandwidth or spacing between adjacent frequency channels (MHz). \\
\texttt{tstart} & Start time of the observation in Modified Julian Date (MJD). \\
\texttt{tsamp} & Sampling time or temporal resolution per time step (seconds). \\
\texttt{ra\_hours} & Right Ascension of the target pointing in hour-angle. \\
\texttt{dec\_degrees} & Declination of the target pointing in degrees. \\
\texttt{telescope\_id} & Identifier code for the observatory telescope used (e.g., VLA, Parkes, MeerKAT). \\
\texttt{num\_timesteps} & Total number of time intervals/integrations present in the entire data block. \\
\texttt{num\_channels} & Total number of fine frequency channels contained in the data block. \\
\texttt{coarse\_channel} & Index of the coarse channel being processed or extracted. \\
\texttt{start\_channel} & Starting index of the fine channel within the broader file structure. \\
\bottomrule
\end{tabularx}
\caption{COSMIC Database Hits table fields}
\label{tab:hit_table_descriptions}
\end{table*}





\clearpage
\newpage


\bibliographystyle{unsrt} 
\bibliography{references}

@ARTICLE{Petigura_2013,
       author = {{Petigura}, Erik A. and {Howard}, Andrew W. and {Marcy}, Geoffrey W.},
        title = "{Prevalence of Earth-size planets orbiting Sun-like stars}",
      journal = {Proceedings of the National Academy of Science},
         year = 2013,
        month = nov,
       volume = {110},
       number = {48},
        pages = {19273-19278},
          doi = {10.1073/pnas.1319909110},
archivePrefix = {arXiv},
       eprint = {1311.6806},
 primaryClass = {astro-ph.EP},
       adsurl = {https://ui.adsabs.harvard.edu/abs/2013PNAS..11019273P}
}

@ARTICLE{Siemion2013,
       author = {{Siemion}, Andrew P.~V. and {Demorest}, Paul and {Korpela}, Eric and {Maddalena}, Ron J. and {Werthimer}, Dan and {Cobb}, Jeff and {Howard}, Andrew W. and {Langston}, Glen and {Lebofsky}, Matt and {Marcy}, Geoffrey W. and {Tarter}, Jill},
        title = "{A 1.1-1.9 GHz SETI Survey of the Kepler Field. I. A Search for Narrow-band Emission from Select Targets}",
      journal = {The Astrophysical Journal},
         year = 2013,
        month = apr,
       volume = {767},
       number = {1},
          eid = {94},
        pages = {94},
          doi = {10.1088/0004-637X/767/1/94},
archivePrefix = {arXiv},
       eprint = {1302.0845},
 primaryClass = {astro-ph.GA},
       adsurl = {https://ui.adsabs.harvard.edu/abs/2013ApJ...767...94S}
}

@ARTICLE{Sardinia,
       author = {{Manunza}, Lorenzo and {Vendrame}, Alice and {Pizzuto}, Luca and {Mulas}, Monica and {Perez}, Karen I. and {Gajjar}, Vishal and {Melis}, Andrea and {Pilia}, Maura and {Perrodin}, Delphine and {Aresu}, Giambattista and {Burgay}, Marta and {Cabras}, Alessandro and {Carboni}, Giuseppe and {Casu}, Silvia and {Coiana}, Tiziana and {Corongiu}, Alessandro and {Croft}, Steve and {Egron}, Elise and {Johnson}, Owen A. and {Ladu}, Adelaide and {Lebofsky}, Matt and {Loi}, Francesca and {MacMahon}, David and {Migoni}, Carlo and {Molinari}, Emilio and {Murgia}, Matteo and {Pellizzoni}, Alberto and {Pisanu}, Tonino and {Poddighe}, Antonio and {Possenti}, Andrea and {Rea}, Erika and {Siemion}, Andrew and {Soletta}, Paolo and {Trudu}, Matteo and {Vacca}, Valentina},
        title = "{The first high frequency technosignature search survey with the Sardinia Radio Telescope}",
      journal = {Acta Astronautica},
         year = 2025,
        month = aug,
       volume = {233},
        pages = {155-167},
          doi = {10.1016/j.actaastro.2025.04.007},
archivePrefix = {arXiv},
       eprint = {2410.09288},
 primaryClass = {astro-ph.HE},
       adsurl = {https://ui.adsabs.harvard.edu/abs/2025AcAau.233..155M}
}

@ARTICLE{Li_2023_SETIDR,
       author = {{Li}, Megan G. and {Sheikh}, Sofia Z. and {Gilbertson}, Christian and {He}, Matthias Y. and {Isaacson}, Howard and {Croft}, Steve and {Sneed}, Evan L.},
        title = "{Developing a Drift Rate Distribution for Technosignature Searches of Exoplanets}",
      journal = {The Astronomical Journal},
         year = 2023,
        month = nov,
       volume = {166},
       number = {5},
          eid = {182},
        pages = {182},
          doi = {10.3847/1538-3881/acf83d},
archivePrefix = {arXiv},
       eprint = {2311.01427},
 primaryClass = {astro-ph.EP},
       adsurl = {https://ui.adsabs.harvard.edu/abs/2023AJ....166..182L}
}

@article{cosmic2023,
  title={COSMIC: An Ethernet-based Commensal, Multimode Digital Backend on the Karl G. Jansky Very Large Array for the Search for Extraterrestrial Intelligence},
  author={Tremblay, Chenoa D and Varghese, Savin Shynu and Hickish, Jack and Demorest, PB and Ng, Cherry and Siemion, Andrew PV and Czech, Daniel and Donnachie, Ross A and Farah, Wael and Gajjar, Vishal and others},
  journal={The Astronomical Journal},
  volume={167},
  number={1},
  pages={35},
  year={2023},
  publisher={IOP Publishing}
}

@INPROCEEDINGS{cosmichighfreq2025AASiposter,
       author = {{Stiegler}, Noah and {Tremblay}, Chenoa and {Cosmic Team}},
        title = "{A High Frequency Technosignature Search with COSMIC and the VLA}",
    booktitle = {American Astronomical Society Meeting Abstracts \#245},
         year = 2025,
       series = {American Astronomical Society Meeting Abstracts},
       volume = {245},
        month = jan,
          eid = {363.01},
        pages = {363.01},
       adsurl = {https://ui.adsabs.harvard.edu/abs/2025AAS...24536301S},
      note = {Accessed August 17th, 2025 at https://aas242-aas.ipostersessions.com/?s=0F-35-E6-14-DC-DB-CC-16-F3-B2-AB-FD-6C-AB-70-54}
}

@article{vlass2025cosmic,
  title={COSMIC’s Large-scale Search for Technosignatures during the VLA Sky Survey: Survey Description and First Results},
  author={Tremblay, Chenoa D and Sofair, Jared and Steffes, Lucy and Myburgh, Talon and Czech, Daniel and Demorest, Paul B and Donnachie, Ross A and Pollak, Alex W and Ruzindana, Mark and APV, Siemion and others},
  journal={The Astronomical Journal},
  volume={169},
  number={3},
  pages={122},
  year={2025},
  publisher={IOP Publishing}
}

@article{hickish2019commensal,
  title={Commensal, multi-user observations with an ethernet-based Jansky Very Large Array},
  author={Hickish, Jack and Beasley, Tony and Bower, Geoff and Burke-Spolaor, Sarah and Croft, Steve and DeBoer, Dave and Demorest, Paul and Diamond, Bill and Gajjar, Vishal and Law, Casey and others},
  journal={arXiv preprint arXiv:1907.05263},
  year={2019}
}

@article{li2022drift,
  title={Drift Rates of Narrowband Signals in Long-term SETI Observations for Exoplanets},
  author={Li, Jian-Kang and Zhao, Hai-Chen and Tao, Zhen-Zhao and Zhang, Tong-Jie and Xiao-Hui, Sun},
  journal={The Astrophysical Journal},
  volume={938},
  number={1},
  pages={1},
  year={2022},
  publisher={IOP Publishing}
}

@article{choza2023breakthrough,
  title={The breakthrough listen search for Intelligent Life: Technosignature search of 97 nearby galaxies},
  author={Choza, Carmen and Bautista, Daniel and Croft, Steve and Siemion, Andrew PV and Brzycki, Bryan and Bhattaram, Krishnakumar and Czech, Daniel and de Pater, Imke and Gajjar, Vishal and Isaacson, Howard and others},
  journal={The Astronomical Journal},
  volume={167},
  number={1},
  pages={10},
  year={2023},
  publisher={IOP Publishing}
}

@article{czech2021breakthrough,
  title={The Breakthrough Listen Search for Intelligent Life: MeerKAT Target Selection},
  author={Czech, Daniel and Isaacson, Howard and Pearce, Logan and Cox, Tyler and Sheikh, Sofia Z and Brzycki, Bryan and Buchner, Sarah and Croft, Steve and DeBoer, David and DeMarines, Julia and others},
  journal={Publications of the Astronomical Society of the Pacific},
  volume={133},
  number={1024},
  pages={064502},
  year={2021},
  publisher={IOP Publishing}
}

@article{mason2025conducting,
  title={Conducting high-frequency radio SETI searches using ALMA},
  author={Mason, Louisa A and Garrett, Michael A and Wandia, Kelvin and Siemion, Andrew PV},
  journal={Monthly Notices of the Royal Astronomical Society},
  volume={536},
  number={3},
  pages={2127--2134},
  year={2025},
  publisher={Oxford University Press}
}

@article{smith2021radio,
  title={A radio technosignature search towards Proxima Centauri resulting in a signal of interest},
  author={Smith, Shane and Price, Danny C and Sheikh, Sofia Z and Czech, Daniel J and Croft, Steve and DeBoer, David and Gajjar, Vishal and Isaacson, Howard and Lacki, Brian C and Lebofsky, Matt and others},
  journal={Nature Astronomy},
  volume={5},
  number={11},
  pages={1148--1152},
  year={2021},
  publisher={Nature Publishing Group UK London}
}

@article{ma2023deep,
  title={A deep-learning search for technosignatures from 820 nearby stars},
  author={Ma, Peter Xiangyuan and Ng, Cherry and Rizk, Leandro and Croft, Steve and Siemion, Andrew PV and Brzycki, Bryan and Czech, Daniel and Drew, Jamie and Gajjar, Vishal and Hoang, John and others},
  journal={Nature Astronomy},
  volume={7},
  number={4},
  pages={492--502},
  year={2023},
  publisher={Nature Publishing Group UK London}
}

@article{cohen1987narrow,
  title={Narrow polarized components in the OH 1612-MHz maser emission from supergiant OH--IR sources},
  author={Cohen, RJ and Downs, G and Emerson, R and Grimm, M and Gulkis, S and Stevens, G and Tarter, J},
  journal={Monthly Notices of The Royal Astronomical Society},
  volume={225},
  number={3},
  pages={491--498},
  year={1987},
  publisher={The Royal Astronomical Society}
}
\end{document}